# An Euler Solver Based on Locally Adaptive Discrete Velocities

B. T. Nadiga
Theoretical Division and CNLS, Los Alamos National Lab.,
Los Alamos, NM 87545

B. T. Nadiga
CNLS, MS B258, Los Alamos National Lab.
Los Alamos, NM 87545, USA

e-mail: balu@lanl.gov
Tel: (505) 667-9466
Fax: (505) 665-2659

# An Euler Solver Based on Locally Adaptive Discrete Velocities

B. T. Nadiga

Theoretical Division and CNLS, Los Alamos National Lab.,

Los Alamos, NM 87545

**Abstract**

A new discrete-velocity model is presented to solve the three-dimensional Euler equations. The velocities in the model are of an adaptive nature—both the origin of the discrete-velocity space and the magnitudes of the discrete-velocities are dependent on the local flow— and are used in a finite volume context. The numerical implementation of the model follows the near-equilibrium flow method of Nadiga and Pullin [1] and results in a scheme which is second order in space (in the smooth regions and between first and second order at discontinuities) and second order in time. (The three-dimensional code is included.) For one choice of the scaling between the magnitude of the discrete-velocities and the local internal energy of the flow, the method reduces to a flux-splitting scheme based on characteristics. As a preliminary exercise, the result of the Sod shock-tube simulation is compared to the exact solution.

## 1. Introduction

A discrete velocity gas is an ensemble of particles with each particle taking on one of a small finite set of allowable velocities [2,3]. Further, the interaction between particles is defined to achieve the desired macro-behavior of the system, which is usually a set of partial differential equations. Such a discretization of the velocity space and definition of the particle interactions (collisions or relaxation or more generally redistribution) also form the basis for the lattice gas and lattice Boltzmann techniques which have been developed over the last eight years [4,5,6 and references therein]. In spite of the inherently compressible nature of discrete-velocity gases, lattice gases and lattice-Boltzmann techniques, their applications in fluid flow modeling have been restricted to the incompressible or very low Mach number regimes [4,5,6 and references therein,7,8,9]. In this article, we present a discrete-velocity gas which for the first time solves the compressible Euler equations without introducing any artifacts of the velocity discretization over a wide range of Mach numbers. This is achieved by letting the discrete-velocities of the model adapt to the local flow conditions [10]. We show also how this scheme can be reduced to a characteristic-based flux-splitting scheme for the Euler equations [11,12].

In the next section, the model is introduced and shown to reproduce the Euler equations. Section 3 gives a step by step numerical evolution of the model on the lines of Nadiga and Pullin [1], and ?? presents a the Sod shock tube simulation [13] with the model. In ?? we show the reduction of the method to characteristic-based flux-splitting and end with some remarks in ??.

## 2. The Discrete-Velocity Model

The discrete-velocity model we consider has 27 velocities ($\mathbf{q}_a, a = 0, 1, \ldots, 26$):

$$\mathbf{q}_a = (q_{ax}, q_{ay}, q_{az}), \quad (q_{ax}, q_{ay}, q_{az}) \in (-q, 0, q). \tag{1}$$

At any given point in space, the model thus has four different speeds $0, q, \sqrt{2}q$, and, $\sqrt{3}q$. There is one velocity with zero speed, six velocities with speed-$q$, twelve velocities with speed-$\sqrt{2}q$, and eight velocities with speed-$\sqrt{3}q$. Note that this model consists of the more familiar two-dimensional nine-velocity model [14,15,16] at three different values of velocity in the $z$-direction $(-q, 0, q)$, thus comprising 27 velocities. Collisions between particles in the model are such that they individually conserve mass, momentum, and energy. In particular, there are two types of collisions which not only conserve mass, momentum and energy but which also change the speeds of the particles involved. (The post-collision pair of speeds is not a mere permutation of the pre-collision speeds.) Describing them in the plane of the collision, the first type involves a speed-$\sqrt{2}q$ particle colliding with a stationary particle to result in two mutually perpendicular speed-$q$ particles. The second type involves a speed-$\sqrt{3}q$ particle colliding with a stationary particle to result again in a pair of particles moving mutually perpendicularly, but now one with speed $q$ and the other with speed $\sqrt{2}q$.

### 2.1 The Stationary Equilibrium Distribution

Thermodynamic equilibrium of the model is defined as the state of detailed balance of all possible collisions. Consider the stationary equilibrium of such a gas: since there are no preferred directions, the particles are identified by their speeds and there are thus four variables $n_0, n_1, n_2$, and $n_3$, where $n_0$ is the probability that a particle is stationary, $n_1$ is a sixth of the probability that a particle has a speed $q$, since there are six speed $q$ velocities in the model, $n_2$ is a twelth of the probability that a particle has speed $\sqrt{2}q$, and $n_3$ is an eigth of the probability that a particle has speed $\sqrt{3}q$. The subscripts in $n_0, \ldots, n_3$ are the square of the speed divided by the unit of speed $q$. Detailed balancing of collisions results in the equilibrium condition

$$\begin{aligned} n_0 n_2 &= n_1^2, \\ n_0 n_3 &= n_1 n_2. \end{aligned} \tag{2}$$

The population densities $(n_0, n_1, n_2,$ and $n_3)$ are further constrained to satisfy the specified hydrodynamic quantities mass $n$ and energy $ne$:

$$\begin{aligned} n &= n_0 + 6n_1 + 12n_2 + 8n_3, \\ ne &= q^2(3n_1 + 12n_2 + 12n_3). \end{aligned} \tag{3}$$

The stationary ($\overline{n_a q_a} = 0$) equilibrium velocity distribution is obtained by the solution of above four equations (2) and (3):

$$
\begin{aligned}
n_0 &= \frac{n}{27}\left(3 - 2\frac{e}{q^2}\right)^3, \\
n_1 &= \frac{n}{27}\left(3 - 2\frac{e}{q^2}\right)^2 \frac{e}{q^2}, \\
n_2 &= \frac{n}{27}\left(3 - 2\frac{e}{q^2}\right)\left(\frac{e}{q^2}\right)^2, \\
n_3 &= \frac{n}{27}\left(\frac{e}{q^2}\right)^3.
\end{aligned}
\tag{4}
$$

## 2.2 The Euler Equations

The equations we seek to model are the Euler equations which describe the compressible inviscid flow of a fluid (here an ideal monatomic gas). Written in a conservative form [17], and with no other body forces, they are

$$
\begin{aligned}
\frac{\partial \rho}{\partial t} + \nabla \cdot \rho \mathbf{u} &= 0, \\
\frac{\partial \rho \mathbf{u}}{\partial t} + \nabla \cdot (p\mathbf{I} + \rho \mathbf{u} \otimes \mathbf{u}) &= 0, \\
\frac{\partial \rho e_t}{\partial t} + \nabla \cdot \{(p + \rho e_t)\mathbf{u}\} &= 0.
\end{aligned}
\tag{5}
$$

where $\otimes$ represents the binary outer product operator, $\mathbf{I}$ is the unit tensor, and $e_t = e + \mathbf{u}^2/2$. The pressure $p$ is given by the ideal gas equation of state for a monatomic gas: $p = 2/3\rho e$.

## 2.3 The Locally Adaptive Discrete Velocities

To represent the above equations as a discrete-velocity gas, we consider the stationary ($\overline{n_a q_a} = 0$) 27-velocity gas discussed above under two simple transformations (see Fig. 1):

- The origin of the discrete-velocity space is translated to $\mathbf{u}(\mathbf{x}, t)$, where $\mathbf{u}(\mathbf{x}, t)$ is the temporally and spatially varying velocity field of the Euler equations.

- The unit of discrete velocity, $q$ is determined locally from the specific internal energy field $e(\mathbf{x}, t)$ of the Euler equations:

$$q(\mathbf{x}, t) = \sqrt{\alpha e(\mathbf{x}, t)} \tag{6}$$

The scaling factor $\alpha$ is a parameter in the model. To insure positivity of the distribution (4), the restriction on $\alpha$ is

$$\frac{2}{3} < \alpha < \infty. \tag{7}$$

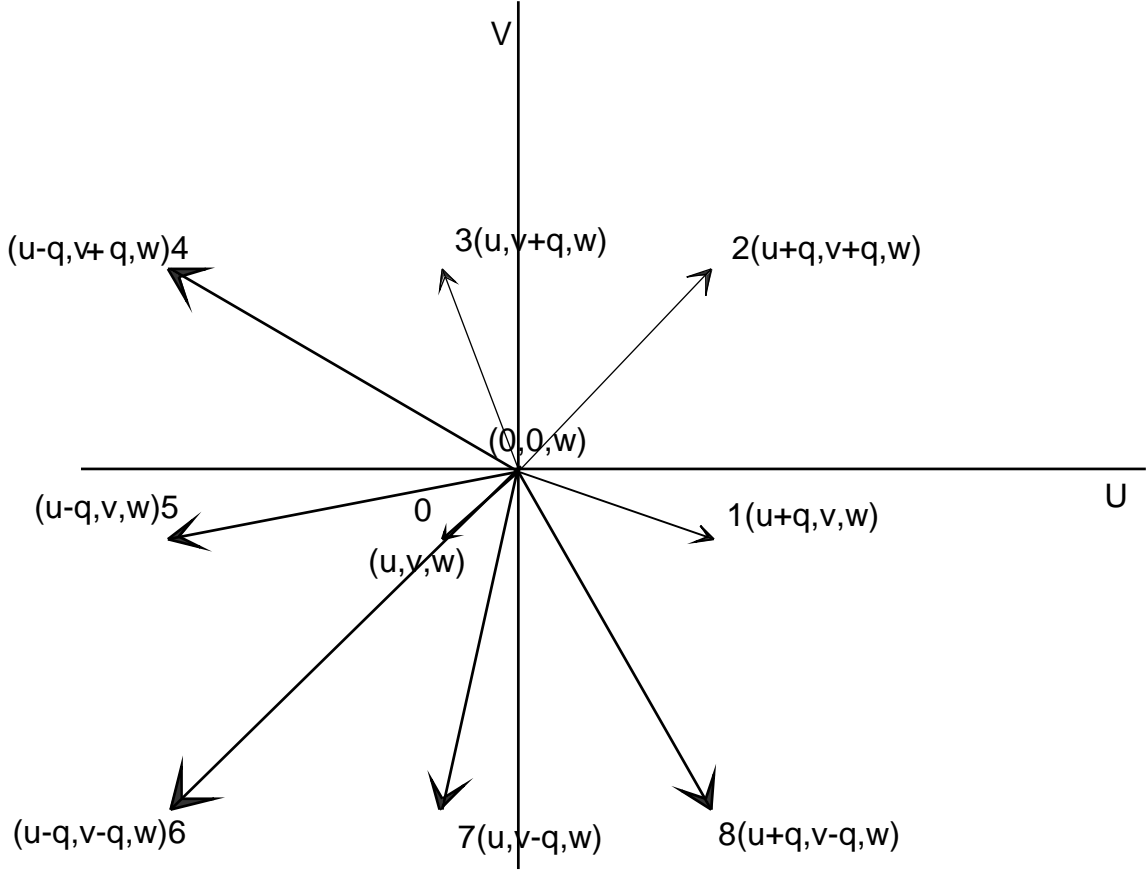

FIG. 1 The origin of the discrete-velocity space is determined by the local flow velocity $\mathbf{u}=(u,v,w)$ of the Euler system and the unit of discrete-velocity $q$ by its specific internal energy. The original nine velocities $(q_{ax}, q_{ay}, 0)$ where $(q_{ax}, q_{ay}) \in (-q, 0, q))$ are shown here after they have adapted themselves to the local macroscopic state. The schematic has been drawn on the $q_{az} = w$ plane, where $w$ is the $z$-component of the macroscopic flow velocity used in the Euler equations. Note that the allowable discrete velocities are now different at each point in space and are different at the same point with time.

Figure 1 shows the original nine velocities $(q_{ax}, q_{ay}, 0; (q_{ax}, q_{ay}) \in (-q, 0, q))$ after they have adapted themselves to the local macroscopic state. The schematic is a projection on to the $q_{az} = w$ plane, where $w$ is the macroscopic velocity in $z$-direction used in the Euler equations (5). With this kind of adaptation, the allowable velocities in the model are now $\mathbf{c}_a(\mathbf{x}, t) = \mathbf{q}_a(\mathbf{x}, t) + \mathbf{u}(\mathbf{x}, t)$, i.e., the allowable discrete-velocities are completely different at each point in space, and are different at the same point in space at different times.

## 2.4 The Equivalence of the model to the Euler Equations

The density and energy of the discrete-velocity gas are set equal to that in (5) and again note that $\overline{n_a q_a}$ has been assumed 0. The equivalence of the discrete-velocity gas system to the system of Euler equations might now be obvious. However, for the sake of completeness, we explain the equivalence. Considering the evolution of the discrete-velocity gas, the (model) Boltzmann equations [3,15]—a statement of the conservation of the number of particles with a particular discrete-velocity—are

$$\frac{\partial n_a}{\partial t} + \mathbf{c}_a \cdot \nabla n_a = Q_a(\mathbf{n},\mathbf{n}), \quad a = 0,\ldots,26, \tag{8}$$

where $Q_a$ is the nonlinear collision operator and the left hand side represents streaming of particles with velocity $\mathbf{c}_a$. The zeroth, first, and second order velocity moments of (8) give respectively, noting that the moments of the collision terms on the right-hand side vanish owing to the mass, momentum, and energy conserving nature of each collision,

$$\frac{\partial n}{\partial t} + \nabla \cdot \overline{n_a \mathbf{c}_a} = 0,$$

$$\frac{\partial \overline{n_a \mathbf{c}_a}}{\partial t} + \nabla \cdot \overline{n_a \mathbf{c}_a \otimes \mathbf{c}_a} = 0, \tag{9}$$

$$\frac{\partial \overline{n_a \frac{\mathbf{c}_a^2}{2}}}{\partial t} + \nabla \cdot \overline{n_a \frac{\mathbf{c}_a^2}{2}\mathbf{c}_a} = 0,$$

where the overbar denotes averaging with respect to the discrete-velocities. From the translation of the origin of the discrete-velocity gas,

$$\begin{aligned}
\overline{n_a \mathbf{c}_a} &= \overline{n_a(\mathbf{u}+\mathbf{q}_a)} = \rho\mathbf{u}, \quad \text{since} \quad \overline{n_a \mathbf{q}_a} = 0. \\
\overline{n_a \mathbf{c}_a \otimes \mathbf{c}_a} &= \overline{n_a(\mathbf{u}+\mathbf{q}_a) \otimes (\mathbf{u}+\mathbf{q}_a)} \\
&= n\mathbf{u}\otimes\mathbf{u} + \overline{n_a \mathbf{q}_a \otimes \mathbf{q}_a} = p\mathbf{I} + \rho\mathbf{u}\otimes\mathbf{u}. \\
\overline{n_a \frac{\mathbf{c}_a^2}{2}\mathbf{c}_a} &= \overline{n_a(\frac{\mathbf{u}^2+\mathbf{q}_a^2}{2} + \mathbf{u}\cdot\mathbf{q}_a)(\mathbf{u}+\mathbf{q}_a)} \\
&= \overline{n_a \frac{\mathbf{q}_a^2}{2}}\mathbf{u} + \rho\frac{\mathbf{u}^2}{2}\mathbf{u} + \overline{n_a(\mathbf{q}_a\cdot\mathbf{u})\mathbf{q}_a} \\
&= \rho e_t \mathbf{u} + \overline{n_a(\mathbf{q}_a \otimes \mathbf{q}_a)\mathbf{u}} = (p+\rho e_t)\mathbf{u}.
\end{aligned} \tag{10}$$

The equivalence is thus complete. Note that in the above equation, since $\overline{n_a \mathbf{q}_a} = 0$, the particles are distributed symmetrically with respect to $\mathbf{q}_a$ and so $\overline{n_a \mathbf{q}_a \otimes \mathbf{q}_a}$ is isotropic and reduces to the pressure $p$.

For convenience, the Euler equations (5) and the moment equations (9) may be rewritten as

$$\frac{\partial \mathbf{f}}{\partial t} + \nabla \cdot \mathbf{G} = 0$$

$$\text{with} \quad \mathbf{f} = \begin{pmatrix} \rho \\ \rho\mathbf{u} \\ \rho e_t \end{pmatrix}, \tag{11}$$

$$\mathbf{G} = [\rho\mathbf{u}, p\mathbf{I} + \rho\mathbf{u}\otimes\mathbf{u}, (p+\rho e_t)\mathbf{u}] = \left[\overline{n_a \mathbf{c}_a}, \overline{n_a \mathbf{c}_a \otimes \mathbf{c}_a}, \overline{n_a \frac{\mathbf{c}_a^2}{2}\mathbf{c}_a}\right].$$

## 3. The Numerical Technique

Hyperbolic systems of conservation laws like the Euler equations (5) in the present case, admit weak solutions in the form of shocks. At these shocks, the gradients of the primary quantities are infinite, and obviously this cannot be represented correctly in a shock-capturing numerical scheme. Shock-capturing schemes are ones in which all grid points are treated in exactly the same fashion irrespective of whether they are inside a shock or outside, and as opposed to shock-tracking schemes which keep track of where the shocks are and treat grid points located in shocks differently from the others. While each have their advantages and disadvantages, used on present day supercomputers, shock-capturing methods are clearly preferable because of their homogenity of computation. Numerically capturing shocks in nondissipative systems like the Euler equations poses the problem of dispersive ripples: a phenomenon in which as a wave steepens, energy flows into the smaller scales and since there is no dissipation, accumulates in the smallest allowed wavelengths—those of the grid-spacing. Thus with the formation of any shock, energy piles up in grid scale oscillations and swamps out scales of interest. A way out of this problem is to dissipate energy at the smallest length scales—the grid-scales—either explicitly or implicitly. Explicit artificial viscosity in general needs adjustments depending on the problem, while physics-based (as opposed to the inviscid idealization represented by the Euler equations) implicit artificial (from the point of view of the Euler equations) viscosity is robust and results in numerical shock widths of the same order as the actual viscous (Navier-Stokes in the present case) shock widths.

We use exactly such an implicit artificial viscosity technique (also called a kinetic numerical scheme owing to the physical kinetic basis of the scheme) which was developed in Nadiga and Pullin [1] for the discrete-velocity framework. A brief description of its usage here follows: For simplicity, consider the computational domain divided into uniform cubical cells, at the centroids of which are stored the cell-averaged values of $(\rho, \rho \mathbf{u}, \text{and } \rho e_t)$. The evolution at each centroid proceeds as follows:

**Step 1**: Calculate the local unit of discrete-velocity using (6) and (7).

**Step 2**: Using (4) and the macroscopic variables of density $\rho$ and specific internal energy $e$, calculate the population densities of the four speeds $n_0, n_1, n_2,$ and $n_3$. Note that in light of restriction (7) and owing to the fact that the stationary distribution function was calculated based on detailed balancing, the above four population densities are necessarily positive.

**Step 3**: Calculate the split fluxes $\mathbf{G}^+$ and $\mathbf{G}^-$ at the centroids using the definitions

$$\mathbf{G}^+ = \begin{pmatrix} \overline{n_a^{c_{ax}>0} c_{ax}} & \overline{n_a^{c_{ax}>0} c_{ax}^2} & \overline{n_a^{c_{ax}>0} c_{ay} c_{ax}} & \overline{n_a^{c_{ax}>0} c_{az} c_{ax}} & \overline{n_a^{c_{ax}>0} \frac{c_a^2}{2} c_{ax}} \\ \overline{n_a^{c_{ay}>0} c_{ay}} & \overline{n_a^{c_{ay}>0} c_{ax} c_{ay}} & \overline{n_a^{c_{ay}>0} c_{ay}^2} & \overline{n_a^{c_{ay}>0} c_{az} c_{ay}} & \overline{n_a^{c_{ay}>0} \frac{c_a^2}{2} c_{ay}} \\ \overline{n_a^{c_{az}>0} c_{az}} & \overline{n_a^{c_{az}>0} c_{ax} c_{az}} & \overline{n_a^{c_{az}>0} c_{ay} c_{az}} & \overline{n_a^{c_{az}>0} c_{az}^2} & \overline{n_a^{c_{az}>0} \frac{c_a^2}{2} c_{az}} \end{pmatrix},$$

$$\mathbf{G}^- = \begin{pmatrix} \overline{n_a^{c_{ax}<0} c_{ax}} & \overline{n_a^{c_{ax}<0} c_{ax}^2} & \overline{n_a^{c_{ax}<0} c_{ay} c_{ax}} & \overline{n_a^{c_{ax}<0} c_{az} c_{ax}} & \overline{n_a^{c_{ax}<0} \frac{c_a^2}{2} c_{ax}} \\ \overline{n_a^{c_{ay}<0} c_{ay}} & \overline{n_a^{c_{ay}<0} c_{ax} c_{ay}} & \overline{n_a^{c_{ay}<0} c_{ay}^2} & \overline{n_a^{c_{ay}<0} c_{az} c_{ay}} & \overline{n_a^{c_{ay}<0} \frac{c_a^2}{2} c_{ay}} \\ \overline{n_a^{c_{az}<0} c_{az}} & \overline{n_a^{c_{az}<0} c_{ax} c_{az}} & \overline{n_a^{c_{az}<0} c_{ay} c_{az}} & \overline{n_a^{c_{az}<0} c_{az}^2} & \overline{n_a^{c_{az}<0} \frac{c_a^2}{2} c_{az}} \end{pmatrix}, \quad (12)$$

where $\overline{n_a^{c_{ax}>0} c_{ax}}$ is the average $\overline{n_a c_{ax}}$ taken only over the discrete velocities $\mathbf{c}_a$ which have a positive $x$-component $c_{ax}$, etc..... For example, when $u > 0$, and $w > 0$, but $v < 0$,

$$\mathbf{G}^+ = \begin{pmatrix} uP + (u+q)Q & u^2P + (u+q)^2Q & uvP + (u+q)vQ & uwP + (u+q)wQ & \overline{n_a^{c_{ax}>0} \frac{c_a^2}{2} c_{ax}} \\ (v+q)Q & (v+q)uQ & (v+q)^2Q & (v+q)wQ & \overline{n_a^{c_{ay}>0} \frac{c_a^2}{2} c_{ay}} \\ wP + (w+q)Q & wuP + (w+q)uQ & wvP + (w+q)vQ & w^2P + (w+q)^2Q & \overline{n_a^{c_{az}>0} \frac{c_a^2}{2} c_{az}} \end{pmatrix},$$

(13)

$$\mathbf{G}^- = \begin{pmatrix} (u-q)Q & (u-q)^2Q & (u-q)vQ & (u-q)wQ & \overline{n_a^{c_{ax}<0} \frac{c_a^2}{2} c_{ax}} \\ vP + (v-q)Q & vuP + (v-q)uQ & v^2P + (v-q)^2Q & vwP + (v-q)wQ & \overline{n_a^{c_{ay}<0} \frac{c_a^2}{2} c_{ay}} \\ (w-q)Q & (w-q)uQ & (w-q)vQ & (w-q)^2Q & \overline{n_a^{c_{az}<0} \frac{c_a^2}{2} c_{az}} \end{pmatrix},$$

where $P = (n_0 + 4n_1 + 4n_2)$ is the density of particles in say the $c_{ax} = u$ plane and $Q = (n_1 + 4n_2 + 4n_3)$ is the density of particles in say the $c_{ax} = u + q$ or $c_{ax} = u - q$ plane. In the above two equations, $\overline{n_a^{c_{ax}>0} \frac{c_a^2}{2} c_{ax}}$, ..., $\overline{n_a^{c_{az}<0} \frac{c_a^2}{2} c_{az}}$ have been left as such for compactness of notation.

**Step 4**: Assuming a linear distribution of the fluxes within the cells in the direction under consideration, interpolate the split fluxes $\mathbf{G}^+$ and $\mathbf{G}^-$ to the cell boundaries:

- interpolate $G_{11}^+$, $G_{12}^+$, $G_{13}^+$, $G_{14}^+$, and $G_{15}^+$ to $(i + \frac{1}{2}, j, k)$,
- interpolate $G_{21}^-$, $G_{22}^-$, $G_{23}^-$, $G_{24}^-$, and $G_{25}^-$ to $(i, j - \frac{1}{2}, k)$, and so on. (Note that the first subscript of $G$ corresponds to the coordinate direction and the second to one of the conserved quantity.)

and apply the minmod limiter to the interpolated fluxes:

$$G_{11}^+(i + \frac{1}{2}, j, k) = G_{11}^+(i, j, k) + \frac{1}{2} \text{minmod}(\Delta_{bck} G_{11}^+(i, j, k), \Delta_{fwd} G_{11}^+(i, j, k)),$$

$$G_{21}^-(i, j - \frac{1}{2}, k) = G_{21}^-(i, j, k) - \frac{1}{2} \text{minmod}(\Delta_{bck} G_{21}^-(i, j, k), \Delta_{fwd} G_{21}^-(i, j, k)), \quad (14)$$

and so on, and where

$$\Delta_{fwd} G_{11}^+(i, j, k) = G_{11}^+(i + 1, j, k) - G_{11}^+(i, j, k)$$

is the first forward difference of $G_{11}^+$ in the $x$-direction at the centroid of cell $(i, j, k)$.

$$\Delta_{bck} G_{21}^-(i, j, k) = G_{21}^-(i, j, k) - G_{21}^-(i, j - 1, k)$$

is the first backward difference of $G_{21}^+$ in the $y$-direction at the centroid of cell $(i, j, k)$ and so on. And minmod is the one dimensional total-variation-diminishing operator as discussed in [18,19]:

$$\text{minmod}(p, q) = \text{sgn}(p) \begin{cases} 0 & \text{if sgn}(p) \neq \text{sgn}(q) \\ \min\{|p|, |q|\} & \text{if sgn}(p) = \text{sgn}(q), \end{cases} \quad (15)$$

with sgn($p$) being the sign of $p$ and $|p|$ the absolute value of $p$.

**Step 5**: Calculate the fluxes at the cell boundaries:
$$\mathbf{G}(\mathbf{i}+\frac{1}{2}) = \mathbf{G}^+(\mathbf{i}+\frac{1}{2}) + \mathbf{G}^-(\mathbf{i}+\frac{1}{2})$$
where $(\mathbf{i}+\frac{1}{2}) = (i+\frac{1}{2}, j, k)$, $(i, j+\frac{1}{2}, k)$, and $(i, j, k+\frac{1}{2})$ in turn.

**Step 6**: Advance the primary variables by one half of the time step to get the midpoint values:
$$\mathbf{f}^{t_0+\Delta t/2} = \mathbf{f}^{t_0} + \frac{\Delta t}{2}\left(\nabla \cdot \mathbf{G}^{t_0}\right), \qquad (16)$$

where e.g. 
$$\nabla \cdot \begin{pmatrix} G_{11} \\ G_{21} \\ G_{31} \end{pmatrix} = \frac{G_{11}(i+\frac{1}{2},j,k) - G_{11}(i-\frac{1}{2},j,k)}{\Delta x} +$$
$$+ \frac{G_{21}(i,j+\frac{1}{2},k) - G_{21}(i,j-\frac{1}{2},k)}{\Delta y} + \frac{G_{31}(i,j,k+\frac{1}{2}) - G_{31}(i,j,k-\frac{1}{2})}{\Delta z}.$$

**Step 7**: Repeat steps 1-4 using the midpoint values $\mathbf{f}^{t_0+\Delta t/2}$ to obtain the fluxes at the midpoint $\mathbf{G}^{t_0+\Delta t/2}$, and take the full time step to obtain the new time level values $\mathbf{f}^{t_1}$:
$$\mathbf{f}^{t_1} = \mathbf{f}^{t_0} + \Delta t\left(\nabla \cdot \mathbf{G}^{t_0+\Delta t/2}\right). \qquad (17)$$

In the above procedure the fluxes were interpolated and limited, instead, the primary quantities **f** could be interpolated and limited. We have done both and the behavior of the two are essentially identical. The time step in the method is limited by the CFL stability criterion
$$\max\left(|U \pm q|\frac{\Delta t}{\Delta X}\right) \leq 1, \quad U \in (u,v,w) \quad \text{and} \quad \Delta X \in (\Delta x, \Delta y, \Delta z). \qquad (18)$$
The code (in Connection Machine Fortran) described by the above step-by-step procedure is included in the Appendix. Notwithstanding the somewhat complicated procedural description above, the simplicity of the resulting code is evident.

## 4. A Numerical Example

The code presented in the Appendix, a Connection Machine Fortran implementation of the second order (in space and time) scheme described in the previous section, was used to calculate the Sod shock-tube problem. The Sod test case has the initial conditions $(\rho_l = 1, u_l = 0, p_l = 1)$ and $(\rho_r = 0.125, u_r = 0, p_r = 0.1)$ corresponding to an initial pressure ratio of 10 and a density ratio of 8. Subscript $l$ denotes the left half and subscript $r$ denotes the right half at time 0. For a monatomic gas, this reduces to the initial conditions $(\rho_l = 1, u_l = 0, e_l = 1.5)$ and $(\rho_r = 0.125, u_r = 0, e_r = 1.2)$. Only 128 points were used for this simulation, at a timestep corresponding to a CFL number of 0.69. (The $\alpha$ in (6) was set at 10/9.) In Fig.2, the exact density, specific internal energy, velocity and pressure profiles (solid lines) are compared to the corresponding profiles (open diamonds) obtained from the discrete velocity simulation. The agreement is good: The shock is typically 3-4 cell-widths and the edges of the rarefaction are not too badly rounded. The spreading of the contact surface is however substantial as with other flux-splitting schemes.

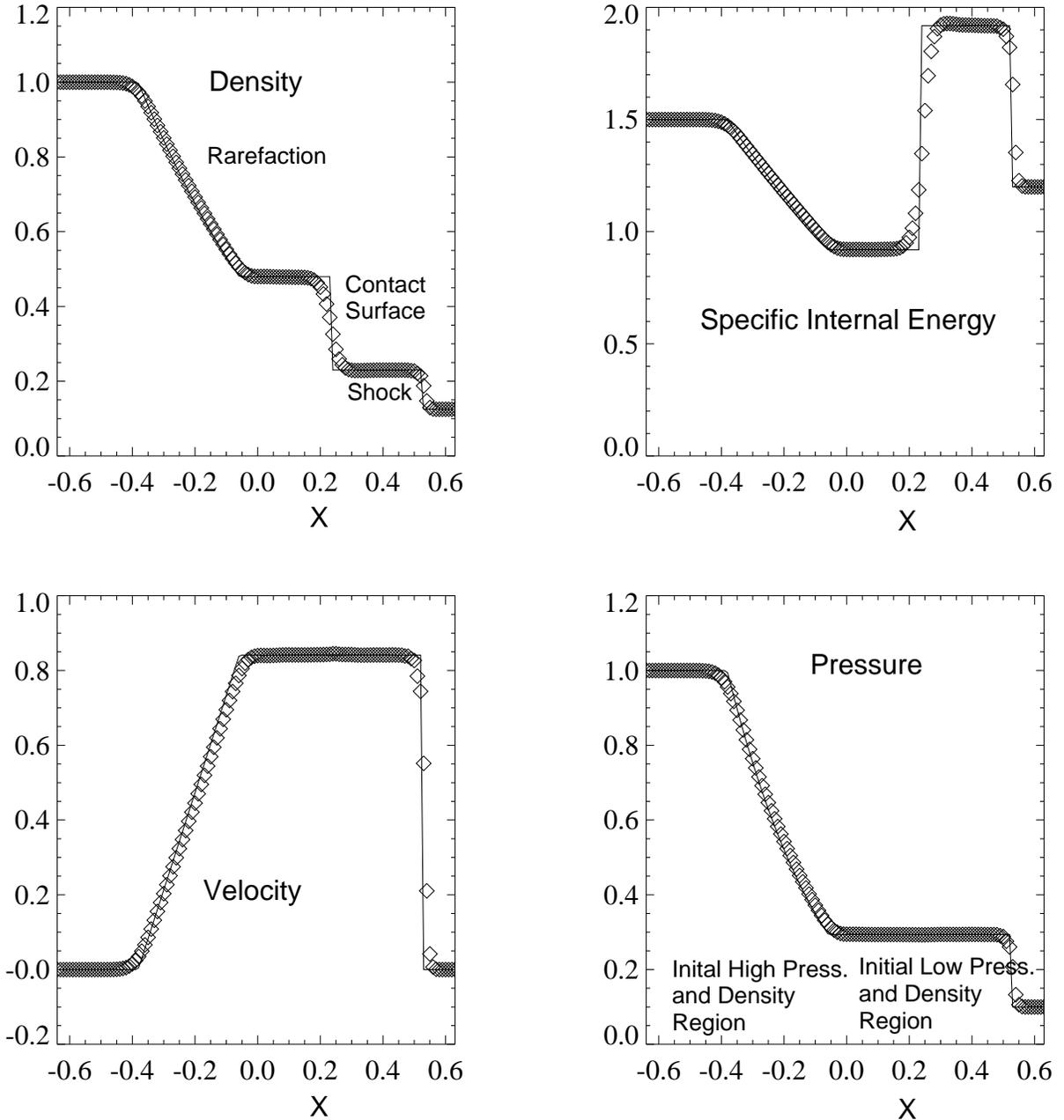

FIG. 2 The Sod shock-tube problem: At a dimensionless time of 0.288 after the diaphragm burst (the diaphragm is initially at x=0), the inital discontinuity in pressure and density (pressure ratio of 10 and a density ratio of 8) is resolved into a right-going shock wave located at about x=0.5 , a left-going rarefaction centered about x=-0.3, and a right-going contact surface at about x=0.2. The solid line is the exact solution and the diamonds are the result of the simulation using the locally adaptive discrete velocities.

## 5. The recovery of Characteristic-based flux-splitting

From the point of view of finite differencing, high-order shock-capturing techniques used for numerically solving inviscid hyperbolic systems of conservation laws are in general based on high-order upwinding. Physically, upwinding can be viewed in two different ways: Firstly, as resulting from a Gudonov methodology of resolving discontinuities in data at cell interfaces using an approximate Riemann solver [11,12]. And secondly, as resulting from the local solution of the collisionless Boltzmann equations [20,21,22]. The present scheme falls into the second category but differs from other such schemes in having a discretized velocity space.

If the unit of discrete-velocity $q(\mathbf{x},t)$ is set equal to the local sound speed $a(\mathbf{x},t)$,

$$q(\mathbf{x},t) = a(\mathbf{x},t) = \sqrt{\gamma RT} = \sqrt{\frac{10}{9}e}, \tag{19}$$

where $\gamma = 5/3$ for a monatomic gas. Next, from the formulae for the split fluxes (13) with $P = 0.4\rho$, $Q = 0.3\rho$, and $q = a$, in the one dimensional case, there are three beams at the three characteristic speeds $u+a, u$, and $u-a$. Now, depending on the signs of $u+a, u$, and $u-a$, the positive fluxes consist of the beams whose speeds are positive and the negative fluxes consist of the beams whose speeds are negative. Thus, the scheme now exactly propagates information along the characteristics of the Euler equations and reduces to a characteristic-based flux-splitting scheme [11,12]. In higher dimensions however, the inherent multi-dimensional (and upwinding) nature of particle motion in the present scheme appears to make it different and needs to be further investigated.

## 6. Conclusion

Wanting to incorporate the simple and elegant way in which discrete-velocity gases (which include the lattice gas and lattice-Boltzmann formulations) combine physics and numerics in the finite-volume techniques for the Euler equations led us to an adaptive discrete velocity model for the Euler equations. The adaptive nature of the discrete velocities (*i.e.* the variable origin and local scaling of the discrete velocities) seems to bridge the gap between the newer discrete-velocity techniques and the more conventional flux-splitting techniques. Based on earlier work, we develop a second order (in space and time) scheme which is very simple and yet robust: it can handle flows over a wide range of Mach numbers accurately and capture shock jumps over 3-4 slab widths with no oscillations. We are presently working on an extension of the method wherein there is a Bhatnagar-Gross-Krook [23] like relaxation of the particle distributions over the timestep $\Delta t$, on the lines of [22]. This will further reduce the diffusive nature of the shocks and the rounding of the corners of rarefaction waves. We wish to emphasis at this point that the new scheme being a finite volume technique, the use of discrete-velocity gases is now possible with arbitrary and irregular spatial meshes. (Though this point was implicit in [1,10], it has sometimes been overlooked.) The limiting scheme used to achieve second order accuracy is however one-dimensional and therefore the second order scheme suffers from all the drawbacks of dimension-split methods. It remains to be investigated if the method can be made genuinely multi-dimensional. We also plan to investigate the extension of the model to solve the Navier-Stokes equations from its present capability of solving the Euler equations.

The full three-dimensional code is presented in the Appendix. It is in CMFortran (a CM extension of Fortran90) for ease of understanding and presentation. A Fortran77 version of the code may be had from the author upon request. This work was supported by DOE in part under the CHAMMP program.

## THE CMF CODE LISTING

### THE MAIN ROUTINE

```
      include '/usr/include/cm/cmssl-cmf.h'
      include 'include.h'
      real,array(5,nx,ny,nz)::fnew,fold
cmf$ layout fnew(:serial,,,),fold(:serial,,,)
c
c INITIALIZE HERE. (Code left out.)
c
      call CMF_cm_array_to_file(22,fnew,ioerr)
      cflav=0
      nav=0
      print *,''
c MAIN LOOP TO DO MIDPOINT INTEGRATION
      do 100 kt=1,nt
        call flux3d(fnew)
        fold=fnew+0.5*dt*dfdt
        call flux3d(fold)
        fnew=fnew+dt*dfdt
        if(mod(kt,ntwrite).eq.0) then
```

```
            print 300,kt*dt,kt,cflav/nav,cfl
            call CMF_cm_array_to_file(22,fnew,ioerr)
         endif
100      continue
300      format('time=',f8.4,'  kt=',i5,'   cflav=',f7.3,'   cfl=',f7.3)
         stop
         end
```

### SUBROUTINE FLUX3D

```
c This subroutine calls flux1d 3 times, once for each dimension.
         subroutine flux3d(fbasic)
         include "include.h"
         real,array(5,nx,ny,nz)::fbasic
cmf$ layout fbasic(:serial,,,)
         dfdt=0.0
         call flux1d(2,3,4,fbasic)
         call flux1d(3,4,2,fbasic)
         call flux1d(4,2,3,fbasic)
         return
         end
```

### SUBROUTINE FLUX1D

```
c For the cell centered at i, this subroutine calculates the
c forward going fluxes at i+1/2 and backward going fluxes at i-1/2
c The limiting procedure is used on the primary quantities.
         subroutine flux1d(iaxis,jax,kax,fbasic)
         include "include.h"
         integer iaxis,jax,kax
         real,array(nx,ny,nz)::tmp11,tmp12,tmp13
         real,array(nx,ny,nz)::rho,u,v,w,e,q
         real,array(5,nx,ny,nz)::fbasic,gxp,gxm,dff
cmf$ layout fbasic(:serial,,,)
cmf$ layout gxp(:serial,,,),gxm(:serial,,,),dff(:serial,,,)
cmf$ layout tmp11(,,),tmp12(,,),tmp13(,,)
cmf$ layout rho(,,),u(,,),v(,,),w(,,),e(,,),q(,,)
```

```
              call minmod(fbasic,iaxis,dff)
              gxp=fbasic+0.5*dff
              gxm=fbasic-0.5*dff
c        Calculate Forward going fluxes at i+1/2 (ru,ruu,rvu,rwu,retu)
              rho=gxp(1,:,:,:)
              u=gxp(iaxis,:,:,:)/rho
              v=gxp(jax,:,:,:)/rho
              w=gxp(kax,:,:,:)/rho
              e=gxp(5,:,:,:)/rho
              e=e-0.5*(u*u+v*v+w*w)
              q=sqrt(alpha*e)
c        Note: correct only for 1D
              cfl=maxval(max(abs(u-q),abs(u+q)))*dt/dxyz(iaxis)
              cflav=cflav+cfl
              nav=nav+1
              e=e/(q**2+1.0e-30)
c Calculate the stationary equilibrium distribution at i+1/2
              tmp=3.-2.*e
              rho=rho/27.
              n0=rho*tmp**3.
              n1=rho*tmp*tmp*e
              n2=rho*tmp*e*e
              n3=rho*e**3.
c Mass
              tmp1(1,:,:,:)=4.*(n2+n3)+n1
              tmp2(1,:,:,:)=4.*(n2+n1)+n0
c X-Momentum
              tmp1(iaxis,:,:,:)=tmp1(1,:,:,:)*(u+q)
              tmp2(iaxis,:,:,:)=tmp2(1,:,:,:)*(u)
c Y-Momentum
              tmp1(jax,:,:,:)=tmp1(1,:,:,:)*v
              tmp2(jax,:,:,:)=tmp2(1,:,:,:)*v
c Z-Momentum
              tmp1(kax,:,:,:)=tmp1(1,:,:,:)*w
              tmp2(kax,:,:,:)=tmp2(1,:,:,:)*w
c Energy
              tmp11=v*v
              tmp12=w*w
              tmp13=2.*q*q
              tmp=2.*(2.*n3+n2)*(tmp11+tmp12+tmp13)+(2.*n2+n1)*(tmp11+tmp12)
              tmp1(5,:,:,:)=0.5*(tmp1(1,:,:,:)*((u+q)*(u+q)+tmp11+tmp12)+tmp)
              tmp2(5,:,:,:)=0.5*(tmp2(1,:,:,:)*(u*u+tmp11+tmp12)+
     >              2.*(2.*n2+n1)*(tmp11+tmp12+tmp13)+(2.*n1+n0)*(tmp11+tmp12))
```

```
              tmp12=(u+abs(u))
              forall(i=1:5)
     >           gxp(i,:,:,:)=(u+q)*tmp1(i,:,:,:)+0.5*(tmp12*tmp2(i,:,:,:))
     c        Calculate Backward going fluxes at i-1/2 (ru,ruu,rvu,rwu,retu)
              rho=gxm(1,:,:,:)
              u=gxm(iaxis,:,:,:)/rho
              v=gxm(jax,:,:,:)/rho
              w=gxm(kax,:,:,:)/rho
              e=gxm(5,:,:,:)/rho
              e=e-0.5*(u*u+v*v+w*w)
              q=sqrt(alpha*e)
              e=e/(q**2+1.0e-30)
              cfl=maxval(max(abs(u-q),abs(u+q)))*dt/dxyz(iaxis)
              cflav=cflav+cfl
              nav=nav+1
              tmp=3.-2.*e
              rho=rho/27.
              n0=rho*tmp**3.
              n1=rho*tmp*tmp*e
              n2=rho*tmp*e*e
              n3=rho*e**3.
              tmp1(1,:,:,:)=4.*(n2+n3)+n1
              tmp2(1,:,:,:)=4.*(n2+n1)+n0
              tmp1(iaxis,:,:,:)=tmp1(1,:,:,:)*(u-q)
              tmp2(iaxis,:,:,:)=tmp2(1,:,:,:)*(u)
              tmp1(jax,:,:,:)=tmp1(1,:,:,:)*v
              tmp2(jax,:,:,:)=tmp2(1,:,:,:)*v
              tmp1(kax,:,:,:)=tmp1(1,:,:,:)*w
              tmp2(kax,:,:,:)=tmp2(1,:,:,:)*w
              tmp11=v*v
              tmp12=w*w
              tmp13=2.*q*q
              tmp=2.*(2.*n3+n2)*(tmp11+tmp12+tmp13)+(2.*n2+n1)*(tmp11+tmp12)
              tmp1(5,:,:,:)=0.5*(tmp1(1,:,:,:)*((u-q)*(u-q)+tmp11+tmp12)+tmp)
              tmp2(5,:,:,:)=0.5*(tmp2(1,:,:,:)*(u*u+tmp11+tmp12)+
     >              2.*(2.*n2+n1)*(tmp11+tmp12+tmp13)+(2.*n1+n0)*(tmp11+tmp12))
              tmp12=(u-abs(u))
              forall(i=1:5)
     >           gxm(i,:,:,:)=(u-q)*tmp1(i,:,:,:)+0.5*(tmp12*tmp2(i,:,:,:))
              tmp1=cshift(gxm,iaxis,1)
              tmp2=cshift(gxp,iaxis,-1)
              dfdt=dfdt+(tmp2-gxp+gxm-tmp1)/dxyz(iaxis)
              return
```

```
        end
```

## SUBROUTINE MINMOD

```
c Given an array y, this routine returns the limited first difference
c using the minmod limiter.
        subroutine minmod(y,j,ydff)
        include "include.h"
        real,array(5,nx,ny,nz)::y,ydff
cmf$ layout y(:serial,,,),ydff(:serial,,,)
        data eps/1.0e-30/
        tmp1=(cshift(y,j, 1)-y)
        tmp2=(y-cshift(y,j,-1))
        ydff=tmp2*tmp1
        ydff=sign(1.,tmp1)*max(0.,ydff)/abs(ydff+eps)
        ydff=ydff*min((abs(tmp1)),(abs(tmp2)))
        return
        end
```

## INCLUDE.H

```
c Note that the number of points in the x direction, nx is
c set to 256 since we are running with periodic boundary conds.
c Only half of that i.e. 128 points are really used.
        integer,parameter::nx=256,ny=16,nz=1
        real,array(nx,ny,nz)::n0,n1,n2,n3,tmp
        real,array(5,nx,ny,nz)::dfdt,tmp1,tmp2
cmf$ layout n0(,,),n1(,,),n2(,,),n3(,,),tmp(,,)
cmf$ layout dfdt(:serial,,,),tmp1(:serial,,,),tmp2(:serial,,,)
        common/scalar/ dt,qurat,alpha,cfl,cflav,nav,dxyz(4)
        common/vector/ n0,n1,n2,n3,tmp,dfdt,tmp1,tmp2
```